\documentclass[12pt]{article} 
\begin{document}   
\newcommand{\todo}[1]{{\em \small {#1}}\marginpar{$\Longleftarrow$}}   
\newcommand{\labell}[1]{\label{#1}\qquad_{#1}} %{\label{#1}} %  
\newcommand{\x}{{\bf x}}
\newcommand{\y}{{\bf y}}
\newcommand{\z}{{\bf z}}
\newcommand{\bp}{{\bf p}}
\newcommand{\A}{{\bf A}}
\newcommand{\B}{{\bf B}}
\newcommand{\p}{\varphi}
\newcommand{\del}{\nabla}
\newcommand{\be}{\begin{equation}}
\newcommand{\ee}{\end{equation}}
\newcommand{\bq}{\begin{eqnarray}}
\newcommand{\eq}{\end{eqnarray}}
\newcommand{\ba}{\begin{eqnarray}}
\newcommand{\ea}{\end{eqnarray}}
\def\r{\nonumber\cr}
\def\hf{\textstyle{1\over2}}
\def\qr{\textstyle{1\over4}}
\def\Sc{Schr\"odinger\,}
\def\sc{Schr\"odinger\,}
\def\>{\rangle}
\def\<{\langle}
\def\-{\rightarrow}
\def\dbd{\partial\over\partial}
\def\tr{{\rm tr}}
\def\pd{\partial}
\def\Det{\mbox{Det}}
\def\Tr{\mbox{Tr}}

\rightline{DCPT-02/57}   
\rightline{hep-th/0207219}
\vskip 1cm

%%                      Title here  
%%  

\begin{center} 
{\Large \bf A Large-D Weyl Invariant String Model}\\
{\Large \bf in Anti-de Sitter Space}
\end{center} 
\vskip 1cm   
  
\renewcommand{\thefootnote}{\fnsymbol{footnote}}   
\centerline{\bf   
 Ian Davies\footnote{I.J.Davies@durham.ac.uk} and Paul Mansfield\footnote{P.R.W.Mansfield@durham.ac.uk}}    
\vskip .5cm   
\centerline{ \it Centre for Particle Theory, Department of  
Mathematical Sciences}   
\centerline{\it University of Durham, South Road, Durham DH1 3LE, U.K.}   
\setcounter{footnote}{0}   
\renewcommand{\thefootnote}{\arabic{footnote}}

%%                      Text starts here  
%% 

\begin{abstract}   

Strings propagating in $AdS_{D+1}$ are made Weyl invariant to leading order
in large $D$ by a ghost-matter coupling that preserves the Poincar\'e symmetry of the boundary.

\end{abstract}

It has been proposed~\cite{55}~\cite{49} that Wilson loops for gauge theories
in flat space with coordinates $X^i$ are described by strings 
in a target space with metric of the form
\be
ds^{2} = d\varphi^{2}+z^{2}(\varphi)\sum_{i=1}^{D}(dX^{i})^{2}\,.\label{met}
\ee
The extra coordinate $\varphi$ is the 
Liouville mode 
and the `warp factor' is taken to be 
$$
z^{2}(\varphi) = \exp\left(\frac{2\varphi}{l}\right)
$$
so that the metric is that of Euclidean $AdS_{D+1}$ space with radius of
curvature $l$. (This metric also occurs in brane-world models and 
in the $AdS$/CFT correspondence.) When the curvature is small on the
string scale, i.e. when  $\alpha^{'}/l^{2}<<1$, the conditions for Weyl
invariance of the string theory can be reliably computed in perturbation theory
and require the presence of  $X^i$-dependent fields such as the dilaton~\cite{18}.
These break the space-time symmetries of the original flat space
and add to the complexity of the model. However one would also like to investigate the regime
where $\alpha^{'}/l^{2}$ cannot be assumed small so that the finite length of the
string plays a significant role. Such situations can arise when the
curvature of the target space geometry is appreciable at the string
scale. In order to probe such situations one must either develop
non-perturbative techniques, or identify some other dimensionless parameter
in the theory which can be taken to be small in the regime of
interest. One could then perform a perturbative computation of the
Weyl anomaly with such a parameter and see how the conventional
conditions for Weyl invariance are changed. We will analyze string propagation in the 
target space of the form (\ref{met}) in an expansion in inverse powers of
the dimension. $D$ plays an analogous role to $N$ in the $O(N)$ nonlinear sigma 
model, which has a well-known large $N$ expansion~\cite{Polybook}. By working to leading order
we obtain a Weyl invariant theory
by tuning $D$ such that $D+1=26$, and by including a coupling between the
Faddeev-Popov ghost sector of the worldsheet theory and the ``matter''
field $\varphi$. Such a coupling can, in part, be interpreted as the
presence of a dilaton field that depends only on $\varphi$. In the
context of Polyakov's conjecture~\cite{54}, this means that the dilaton no
longer breaks Poincar\'e and rotational invariance in the boundary
gauge theory, unlike the dilaton field that one requires in the
conventional small $\alpha^{'}/l^{2}$ Weyl anomaly analysis. (Coupling 
matter and ghost sectors has been studied in different contexts in~\cite{edwitten}~\cite{Siegel}.)

           To pursue
the analogy with the $O(N)$ nonlinear sigma model we write the string partition function in terms of coordinates  
$W^{i} = zX^{i}$ so that

$$
Z = \int\mathcal{D}g\:\mathcal{D}z\:\mathcal{D}W\:\exp\left(-S[g,W,z]\right)
$$
with
$$
S[g,W,z] = \frac{1}{4\pi}\int
d^{2}\xi\sqrt{g}\left[g^{ab}l^{2}\frac{\pd_{a}z\pd_{b}z}{z^{2}}+W^{i}\left(\Delta-\frac{1}{z}\Delta z
\right)W^{i}\right] 
$$
$\Delta$ is the usual covariant worldsheet Laplacian, and we work in
units such that $\alpha^{'}$ is equal to one. The indices $a,b$ label the two
worldsheet coordinates $\xi_{1},\xi_{2}$. We 
ignore world sheet boundary effects. Performing the Gaussian $W$-integral
generates the functional determinant of the operator
$$
\Gamma = \left(\Delta-\frac{1}{z}\Delta z \right)
$$
The explicit
dependence of $\mbox{Det}^{-D/2}(\Gamma)$ on the metric scale
$\phi(\xi)$ in the conformal gauge $g_{ab}(\xi) =
e^{\phi(\xi)}\delta_{ab}$ can be found using heat kernel techniques,
\bq
\mbox{Det}^{-D/2}(\Gamma) &=& \exp\left(\int
d^{2}\xi\left[\frac{D}{96\pi}(\pd_{a}\phi)^{2}-\frac{D}{8\pi}\frac{\phi}{z}\pd_{a}^{2}
z +\cdots
\right] \right)\nonumber \\
&\equiv&\exp\left(\int
d^{2}\xi\left[A+B+\cdots
\right] \right)\label{bigresult}
\eq
where $(\cdots)$ represents $\phi$-independent terms that we will
consider in a moment. The first term in (\ref{bigresult}), $A$, is the usual contribution to the Weyl anomaly from
$D$ free bosons~\cite{13}, and can be combined with the standard Faddeev-Popov gauge
fixing determinant to give a factor of
$$
\exp\left(\frac{D-26}{96\pi}\int
d^{2}\xi\left[(\pd_{a}\phi)^{2}+\lambda e^{\phi}\right]\right) = \exp\left(\frac{D-26}{96\pi}S_{L}\right)
$$
where $S_{L}$ is the Liouville action for $\phi(\xi)$ in conformal
gauge. (The integral over $z$ will produce a further contribution to this.)

In critical string theory the overall $\phi$ dependence of $Z$ must be cancelled
for consistency. This is also true in non-critical string theory 
since the split of the scale of the metric into the dynamical $\varphi$ and the fiducial
$\phi$ is arbitrary~\cite{David}~\cite{Distler}. Clearly terms proportional to $S_L$ can be cancelled
by tuning $D$ (to a large value).

The second term in (\ref{bigresult}), $B$,  is more problematic. In the usual small 
$\alpha^{'}/l^{2}$
calculation similar terms would appear from expanding the action in Gaussian normal coordinates
about a point, so 
$$
G_{\mu\nu}(\hat{X})\partial_a \hat X^\mu\partial_b \hat X^\nu =\delta_{\mu\nu}\partial_a x^\mu\partial_b x^\nu 
-\frac{1}{3}R_{\mu\lambda\nu\kappa}(C)x^\lambda x^\kappa\partial_a
x^\mu\partial_b x^\nu +\cdots
$$
where $\hat X^\mu=C^\mu+x^\mu$. Contracting $x^\lambda $ with $x^\kappa$ produces a
term proportional to $\phi R_{\mu\nu}\partial_a x^\mu\partial_b x^\nu $. A partial integration would turn this into
something resembling $B$ when the latter is expanded about a constant value of $z$.
Further contraction of
$\partial_a x^\mu$ with $\partial_b x^\nu $ will contribute to the dilaton beta function
at sub-leading order. The term in (\ref{bigresult}) is, by contrast, of leading order in $D$.
As we will see the $z$-propagator is different at leading order in the large-$D$
expansion, consequently $B$ does not contribute to the dilaton beta-function,
but it does make the $z$-propagator depend on $\phi$. This does not happen in the small $\alpha^{'}/l^{2}$
calculation because $x\partial^2 x$ is proportional to
the equation of motion. So to make the $z$-propagator independent of
$\phi$ at leading order in $D$ we need to cancel $B$.
This can be done without breaking the symmetries of the $X$ space if we 
couple $z$ to the ghosts.

Firstly  ``bosonize'' the Faddeev-Popov ghosts by
writing the Faddeev-Popov determinant as
$$
\mbox{Det}^{'}_{FP} = \exp\left(\frac{26}{96\pi}S_{L}\right) \sim \int\mathcal{D}\psi\exp\left(-\frac{1}{96\pi}\int
d^{2}\xi\sqrt{g}\left[\psi\Delta\psi+2\beta R^{(2)}\psi\right]\right)
$$
where $R^{(2)}$ is the scalar curvature of $g_{ab}$, the prime
denotes omission of the zero modes, and $\beta^{2}=-25$. Adding
$$
S_{ct} = \frac{D}{8\pi\beta}\int d^{2}\xi\sqrt{g}\:\frac{\psi}{z}\Delta z
$$
to the action
cancels $B$ in equation
(\ref{bigresult}) on integration over $\psi(\xi)$.  Integration by parts
shows that this counterterm can be interpreted as a dilaton field~\cite{Banks} of
the form $\Phi(z) \sim \ln(z)$, as well as a coupling of the target
space metric to the ghost $\psi(\xi)$ of the form $ds_{gh}^{2}\sim
(\psi/z) dz^{2}$.

In order to complete the calculation of the partition function we
must also determine the $z$-dependence of $\mbox{Det}^{-D/2}(\Gamma)$. This is facilitated by the  $1/D$ expansion. 
We write the
$z(\xi)$ field as
$$
z(\xi)=z_{0}+\frac{\bar{z}(\xi)}{\sqrt{D}}
$$
where $z_{0}$ is a constant on the world-sheet and $\bar{z}(\xi)$ are quantum fluctuations
about $z_{0}$ to be  integrated over. To obtain an effective action for the
$\bar{z}(\xi)$ fluctuations, the determinant of $\Gamma$ is expanded in powers of $\bar{z}(\xi)$ retaining terms up to
$O(1)$ in $D$ . This is most easily done in momentum
space and gives the leading order effective theory
\be
Z \sim \exp\left(\frac{D-26}{96\pi}S_{L}\right)\times
\int\mathcal{D}\bar{z}\:\exp\left(\frac{1}{8\pi}\int\frac{d^{2}p}{(2\pi)^{2}}\frac{1}{z_{0}^{2}}\bar{z}(p)\bar{z}(-p)p^{2}\ln\left(\frac{p^{2}}{\Lambda^{2}}\right)\right) \label{zbaract4}
\ee
(note that the Fourier transform of the $\bar{z}$-propagator is not
the usual $1/p^{2}$, as was mentioned earlier. It is the improved ultra-violet behaviour that is responsible for $B$ not contributing to the dilaton
beta-function in next to leading order). $\Lambda^{2}$ is a
finite overall scale. We need to integrate
over the fluctuations to obtain the final form of the Weyl anomaly,
and this will involve computing the $\phi$-dependence of the
determinant of the operator whose Fourier representation is given in equation
(\ref{zbaract4}). Again, heat kernel techniques and the fact
that this operator derives from $\Gamma$ (whose $\phi$-dependence is
readily found to be $\delta_{\phi}\Gamma = -\delta\phi\Gamma$) gives
a contribution to
the Weyl anomaly equal to that of a single free boson. Hence, the final Weyl anomaly
is found to be
$$
Z \sim \exp\left(\frac{D+1-26}{96\pi}S_{L}\right)
$$
and so $Z$ is Weyl invariant when $D+1=26$, as usual.

In conclusion we have found that to leading order in a large-$D$ expansion it is possible to construct a Weyl
invariant string theory in Euclidean $AdS_{D+1}$ space without breaking the symmetries associated
with the boundary. This is achieved by coupling the $z$ coordinate
to the ghosts, resulting in part in a $z$-dependent dilaton.
The $1/D$ 
expansion promotes to leading order terms that are usually negligible for
small $\alpha^{'}/l^{2}$. The question of whether the ghost-matter coupling  
is necessary to achieve Weyl invariance, rather than just sufficient, is one which warrants
further investigation, as is the question of next to leading order effects.

\vskip.5in

\centerline{\bf Acknowledgements}
\medskip 

Ian Davies would like to thank James Gregory, Ken Lovis, Tony Padilla and
David Page for helpful discussions. Ian Davies is supported by a PPARC
research studentship.

\bibliographystyle{utphys}  
   
\bibliography{ref}  
\end{document}